\documentclass[epj]{svjour}

\usepackage{graphicx}
\usepackage{dcolumn}
\usepackage{bm}

\begin{document}

\title{Statistical analysis of the transmission based on the DMPK equation: An application to Pb nano-contacts\\}

\author{V\'ictor A. Gopar \inst{1,2}}
\institute{$^1$Instituto de Biocomputaci\'on y F{\'i}sica de  Sistemas Complejos,
   Corona de Arag\'on 42, 50009 Zaragoza, Spain \\
$^2$Departamento de F\'isica Te\'orica, Facultad de Ciencias, Universidad de Zaragoza, Pedro Cerbuna 12, 50009 Zaragoza, Spain}

\abstract{
The density of the transmission eigenvalues of Pb nano-contacts has been estimated recently 
in mechanically controllable break-junction experiments. Motivated by these experimental analyses, here we study the evolution of the density of the transmission eigenvalues with the disorder strength  and the number of  channels supported by the ballistic constriction of a quantum point contact in the framework of the  Dorokhov-Mello-Pereyra-Kumar  equation. We find that the transmission density evolves  rapidly into  
the density in the diffusive metallic regime as the number of channels $N_c$ of the constriction increase. Therefore, the transmission density distribution for a few $N_c$ channels comes close to the known bimodal density distribution in the metallic limit. This is in agreement with the experimental statistical-studies in Pb nano-contacts. For the two analyzed cases,  we show that the experimental densities are seen to be well described by the corresponding theoretical results.
}

\titlerunning{Statistical analysis of the transmission based on the DMPK equation: An application to nano-contacts}
\authorrunning{V. A. Gopar}
\maketitle
\section{Introduction}
Since first transport experiments in two-dimensional electron gases confined in semiconductor heterostructures  with a constriction comparable to the Fermi wave length --quantum point contacts-- where conductance quantization were reported   
\cite{wees,wharam}, much attention has been given to describe  theoretically the quantum electron transport in these kind of systems. The dimensions of those structures are small enough that quantum interference effects can be observed in transport measurements, i.e., the typical size of the systems is smaller than the phase-coherent length. More recently, scanning tunneling microscope
and mechanically controllable break-junction (MCBJ) techniques have made possible the fabrication of structures at the atomic scale. We refer to the review article \cite{agrait} for a detailed description of the progresses on those experimental techniques. In particular, in MCBJ experiments a notched wire is elongated with high mechanical precision and stability; thus, during this process atomic-scale contacts, or quantum point contacts, are formed between two wide electrodes. 

From the theoretical side, several aspects of the transport in quantum point contacts have been investigated and an extensive literature already exists. We  refer the reader to  \cite{agrait} and  \cite{carlo-review}  for a review
of the topic. Here we mention some examples of the problems addressed in the literature where random configuration of impurities in the system is an important ingredient and therefore a statistical study of the transport is relevant. For instance, the suppression of the conductance fluctuations due to  disordered reservoirs attached to a ballistic constriction was studied early in the nineties \cite{maslov,carlo-melsen}. Statistical properties of the conductance, resistance, and shot noise in disordered wires with a constriction were studied in Ref. \cite{carlo-melsen}. More recently, effects of impurities around a quantum point contact on the statistics of the transmission eigenvalues have been investigated \cite {campagnano}.  In single-atom contacts, it has been conjectured and experimentally verified that the number of conducting channels  is determined by the valence orbitals of the constriction atom  \cite{agrait,elke}. 

Lately, several aspects of the electron transport in quantum point contacts have been experimentally analyzed using MCBJ techniques  \cite{agrait,smit,riquelme}.   For example, statistical properties of the transport like the density of the transmission eigenvalues have been estimated for Pb nano-contacts \cite{riquelme,memory}; in these experiments, it has been pointed out that the transmission density for contacts with a few open channels  is unexpectedly similar to the known bimodal density distribution in the diffusive metallic regime. We recall that in the metallic limit a  large number of channels are assumed to contribute to the transport.

\begin{figure}
\begin{center}
\includegraphics[width=0.8\columnwidth]{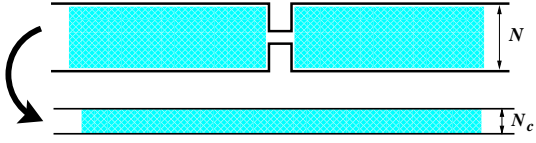}
\end{center}
\caption{\label{fig1} The upper sketch represents  a quantum wire with mean free path $l$ and a ballistic constriction in the middle. The wire can support $N$ channels. As explained in the text, the DMPK equation for such kind of wire can be mapped onto a DMPK equation describing a wire (as the one represented in the lower sketch) without the constriction, mean free path $l'$, and number of channels $N_c$.
}
\end{figure}

Motivated by recent break-junctions experiments where some  
statistical properties of the transmission eigenvalues have been estimated for Pb nano-contacts \cite{riquelme,memory}, in this paper we calculate the density of the transmission eigenvalues within the framework of the  Dorokhov-Mello-Pereyra-Kumar (DMPK) equation \cite{pier-book,dorokhov}. 
In order to apply DMPK ideas to those break-junctions experiments, we assume that the constriction, or neck, formed during the elongation processes of the MBCJ device is smaller or of the order of the mean free path. This nowadays is achieved experimentally; in fact, the narrowest part of a MBCJ can be of molecular-size scale. We will consider that the random character of the electronic transport is due to the disordered wide regions at both sides of the contact. Finally, we suppose that the neck and the wide disordered region of the MBCJ device have a step geometry (as in Fig. \ref{fig1}, upper sketch). 
Thus, our calculations are based on the  
the solution of the DMPK equation given in Refs. \cite{rejaei,go-mu-wo,mu-wo-go} and a mapping of the transport problem between constricted and unconstricted wire geometries introduced in Ref. \cite{carlo-melsen}.  We will show that in agreement with the results reported in Ref. \cite{riquelme}, our theoretical calculations show that the transmission density
can be close to that one in the  diffusive metallic limit, even when the ballistic constriction of the point contact  supports a few  channels.

In subsection \ref{subsection_dmpk}, we introduce the DMPK equation and its solution for a single and multiple open channels. In the same subsection,  
we present the above-mentioned mapping of the problem of a wire with a constriction to that one without such constriction. In  Sec. \ref{section_density},  we introduce the functions of the transmission density that  we study in this work. Our analysis of the density  start in subsection \ref{subsection_one} with the simplest case of a point contact with a ballistic constriction supporting one channel. In subsection \ref{subsection_several}, we analyze the case when several open channels are supported in the ballistic constriction. Also we compare some of the results for the transmission density reported in Ref. \cite{riquelme} with our calculations. Finally, we give a summary of our study in Section \ref{section_conclusions}.

\subsection{\label{subsection_dmpk}The DMPK equation}

A scaling equation for the transmission eigenvalues $\{\tau_i\}$   of a quasi-one-dimensional disordered system, or quantum wire, has been developed in the past  \cite{pier-book,dorokhov}. Within this framework, the evolution of the distribution of the transmission
eigenvalues $p(\{\tau_i\})$ [or equivalently $p(\{\lambda_i\})$, where $\lambda_i=(1-\tau_i)/\tau_i$] with the length $L$ of the quantum wire is given by the Fokker-Planck equation 
\begin{eqnarray}
\label{dmpk}
l\frac{\partial p(\lambda)}{\partial L}&=&\frac{2}{N+1}
\frac{1}{J({\lambda})} \times \nonumber \\ 
& & \sum_a\frac{\partial}
{\partial \lambda_a}\left[\lambda_a(1+\lambda_a)
J(\lambda)\frac{\partial p(\lambda)}
{\partial \lambda_a}\right],
\end{eqnarray}
known as DMPK equation, where $N$ is the number of channels, or transverse modes, and $l$ is the mean free path, while $J(\lambda)=\prod_{i<j}^{N}|\lambda_i -\lambda_j|$. In Eq. (\ref{dmpk}), 
we have assumed a system with time-reversal symmetry ($\beta=1$), although systems without  time-reversal invariance ($\beta=2$) or with broken spin-rotation symmetry ($\beta=4$) can be considered. Also, in the derivation of Eq. (\ref{dmpk}), it was assumed that the width of the wire is constant and smaller than its length  $L$, in order to neglect
diffusion in the transverse direction.

For one channel ($N=1$), the solution of Eq. (\ref{dmpk}) is given by  \cite{carlo-review,abrikosov}
\begin{equation}
\label{poflambda1}
p(\lambda)=\frac{1}{\sqrt{2\pi}}\Big(\frac{1}{s}\Big)^{\frac{3}{2}}
{\rm e}^{-s/4}\int_{y_0}^{\infty}dy\frac{y{\rm e}^{-y^2/4s}}
{\sqrt{\cosh{y}-1-2\lambda}},
\end{equation}
where $y_0={\rm arccosh}{(1+2\lambda)}$ and $s=L/l$ is the length of the system $L$
in units of the mean free path $l$.

The relation between the random variables $\lambda$ (or $\tau$) and the electronic transport in the wire is given through the Landauer formula
\begin{equation}
g=\sum_{i}^{N} \tau_i =\sum_{i}^{N}\frac{1}{1+\lambda_i} ,
\end{equation}
where $g$ is the dimensionless conductance. Thus, the statistical properties of $g$ are governed by that ones of the transmission eigenvalues.

The solution of the DMPK equation is known also for the general case of multiple channels
and for any symmetry ($\beta=1,2,4$) \cite{rejaei,caselle}. The general solution, unfortunately, is quite complicated, but it simplifies 
in the metallic ($1 \ll s \ll N$) and insulating ($L>>Nl$ ) regimes. In these two opposite regimes of transport, the solution of the DMPK equation can be written in the following form
\begin{equation}
\label{poflambda}
p(\lambda)=\frac{1}{Z}\exp[- H(\lambda)],
\end{equation}
where $Z$ is a normalizing factor, and $H(\lambda)$ is given by 
\begin{equation}
\label{H}
H(\lambda)=\sum_{i<j}^{N}u(\lambda_i,\lambda_j)
+\sum_{i}^{N}V(\lambda_i).
\end{equation}
Following Refs. \cite{go-mu-wo} and \cite{mu-wo-go}, we show below that we can use this simplified solution, Eqs. (\ref{poflambda}) and (\ref{H}),  in our analysis.  
Making the change of variables $\lambda_i=\sinh^2 x_i $, the functions 
$u(x_i,x_j)$ and $V(x_i)$ in the diffusive metallic regime are 
given by
\begin{eqnarray}
\label{uandVmetal}
&&u(x_i,x_j)=-\frac{1}{2} \ln \vert \sinh^2 x_i-\sinh^2 x_j\vert
-\frac{1}{2} \ln \vert x^2_i - x^2_j\vert,  \nonumber\\
&&V(x_i)=\frac{(N+1)}{2s}x^2_i-\frac{1}{2} \ln \vert
x_i \sinh 2x_i\vert, 
\end{eqnarray}
while in the insulating regime
\begin{eqnarray}
\label{uandVinsulator}
&&u_{ins}(x_i,x_j)=-\frac{1}{2} \ln \vert \sinh^2 x_i-\sinh^2 x_j\vert
-\ln \vert x^2_i - x^2_j\vert,  \nonumber\\
&&V_{ins}(x_i)=\frac{(N+1)}{2s}x^2_i-\frac{1}{2} \ln \vert
x_i \sinh 2x_i\vert -\frac{1}{2}\ln x_i
\end{eqnarray}
In writing Eqs. (\ref{uandVmetal}) and (\ref{uandVinsulator}) we have assumed the presence of time-reversal invariance in the system ($\beta=1$). We note that  Eq. (\ref{uandVinsulator}) differs from  Eq.  (\ref{uandVmetal}) 
only in the $\ln |x_i^2- x_j^2|$ and $\ln x_i$ terms. We remark 
that in the metallic regime a large number of channels, or $x_i$ variables, contribute to the conductance, whereas in the insulating regime, the main contribution to the conductance comes from one (the smallest one) $x_i$. On the other hand, in the insulating regime, the average conductance is very small and $x_i >>1$; in this limit, $\ln |x_i^2- x_j^2|$ and $\ln x_i$ are negligible compared to the other terms in the insulating case Eq. (\ref{uandVinsulator}). Since these two terms are the only ones that differ from the metallic solution, we thus assume that the metallic solution can be used as an approximated solution to DMPK equation valid beyond the metallic limit and for a finite number of channels.

Finally, we recall that in the diffusive metallic regime, the density of the transmission eigenvalues  defined by the average $\langle \sum_i \delta(\tau -\tau_i)\rangle$ is  known to follows the  bimodal distribution \cite{carlo-review,pier-pichard}:
\begin{equation}
\label{rho-metal}
 \rho_{\mathrm{met}}(\tau) =\frac{\left< g \right>}{2\tau \sqrt{1-\tau}} , 
\end{equation}
with $\langle g \rangle = Nl/L$.
\subsection{Correspondence between constricted and unconstricted wire geometries}

As mentioned in the  previous subsection, the DMPK equation was derived for a wire
geometry, i.e., a quasi-one-dimensional system of length $L$ and width $W$. On the other hand, however, in MCBJ techniques a notched wire is bended to produce a small constriction between two wide electrodes. This constriction can be of the order of the mean free path of the sample, i.e., a ballistic constriction. The problem of transport through a quasi-one-dimensional wire with a  constriction can be mapped on to a problem of a quantum wire without the  constriction, as it was shown by Beenakker and Melsen \cite{carlo-melsen}: it turns out that the DMPK equation for a wire with a ballistic constriction with mean free path $l$ and $N$ number of channels 
(Fig. \ref{fig1}, upper wire) is equivalent to that one for a  wire geometry (without the constriction, Fig. \ref{fig1}, lower wire) with mean free path $l'$ given by 
\begin{equation}
l'=l (N+1)/(N_c+1) ,
\end{equation}
where $N_c$ is the number of channels determined by the ballistic constriction in the wire. We note that a step-constriction geometry is assumed in the model. This mapping  allows us to use the known results 
for unconstricted wire geometries, in particular the solution Eq. (\ref{poflambda}-\ref{uandVmetal}), to study the statistical properties of transport in quantum point contacts; we will be especially interested in functions of the transmission density introduced in  Ref. \cite{riquelme}.

\section{\label{section_density}Density of the transmission eigenvalues}
\begin{figure}
\begin{center}
\includegraphics[width=1\columnwidth]{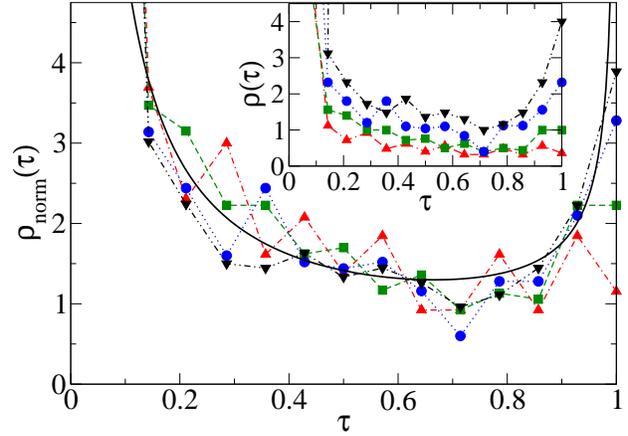}
\end{center}
\caption{(Color on-line) \label{fig2}Density of the transmission eigenvalues as defined in Eq. (\ref{rhonorm}) for Pb nano-contacts with conductances (in units of $e^2/h$) $g =$ 4 (solid upward triangles), 6 (solid squares), 10 (solid circles), and 15  (solid downward triangles). The solid (black) line is the bimodal distribution given by Eq. 
(\ref{rho-metal}). The inset shows the density $\rho(\tau)$ for the same values of the conductances of  the main frame. The data points for $\rho_{\mathrm{norm}}(\tau)$ and $\rho(\tau)$ have been obtained from Refs. \cite{riquelme} and \cite{memory}.}
\end{figure}

The density of the transmission eigenvalues $\rho (\tau)$ were estimated in Ref. \cite{riquelme} by analyzing several samples of Pb nano-contacts. To be specific, the functions $\rho_{\mathrm{norm}}(\tau)$ and $q(\tau)$ defined as
\begin{equation}
\label{rhonorm}
 \rho_{\mathrm{norm}}(\tau)=\frac{1}{\left< g \right>}\rho(\tau) ,
\end{equation}
and 
\begin{equation}
\label{qoftau}
 q(\tau)=\frac{1}{\left< g \right>}\int_0^\tau d\tau' \tau'\rho(\tau') ,
\end{equation}
were introduced by Riquelme et. al. \cite{riquelme,memory}. We find convenient and motivating to reproduce here some of the results of Refs. \cite{riquelme} and  \cite{memory}. In figure 2, we show  $\rho_\mathrm{norm}(\tau)$ and $\rho(\tau)$ for Pb contacts with conductances ranging from 4 to 15 (in units of the conductance quantum $2e^2/h$ ) \cite{data}. We can observe in Fig. \ref{fig2} that $\rho_\mathrm{norm}(\tau)$ 
comes close to the metallic limit result [solid line, Eq. (\ref{rho-metal})], even for contacts with small conductances. This was the unexpected result pointed out by the authors of Ref. \cite{riquelme}.

Coming back to the theoretical model, according to the mapping introduced above and assuming that $N_c$ is the number of channels determined by the constriction in the wire, the transmission density is given by 
\begin{equation}
\label{rhooftau}
\rho(\tau)=\left< \sum_i^{N_c} \delta \left( \tau - \tau_i \right) \right> .
\end{equation}

\subsection{\label{subsection_one}One-channel case}
We start with the simplest case of a quantum point contact with one open channel ($N_c=1$).
It is instructive to start with this case, although no experimental data are available to compare to.  
Using that $\tau=1/(1+\lambda)$ in Eq. (\ref{poflambda1}),  we immediately obtain for the density:
\begin{equation}
\label{rhotauNc1}
\rho(\tau)=\frac{1}{\sqrt{2\pi}}\Big(\frac{\xi}{2 L}\Big)^{\frac{3}{2}}
\frac{{\rm e}^{-L/(2\xi)}}{\tau^2}\int_{y_0}^{\infty}dy\frac{y{\rm e}^
{-\xi y^2/8L}}
{\sqrt{\cosh{y}+1-2/\tau}},
\end{equation}
where $y_0={\rm arccosh}{(2/\tau-1)}$. We also have defined the parameter $\xi=(N+1)l$
in Eq. (\ref{rhotauNc1}).

\begin{figure}
\begin{center}
\includegraphics[width=1\columnwidth]{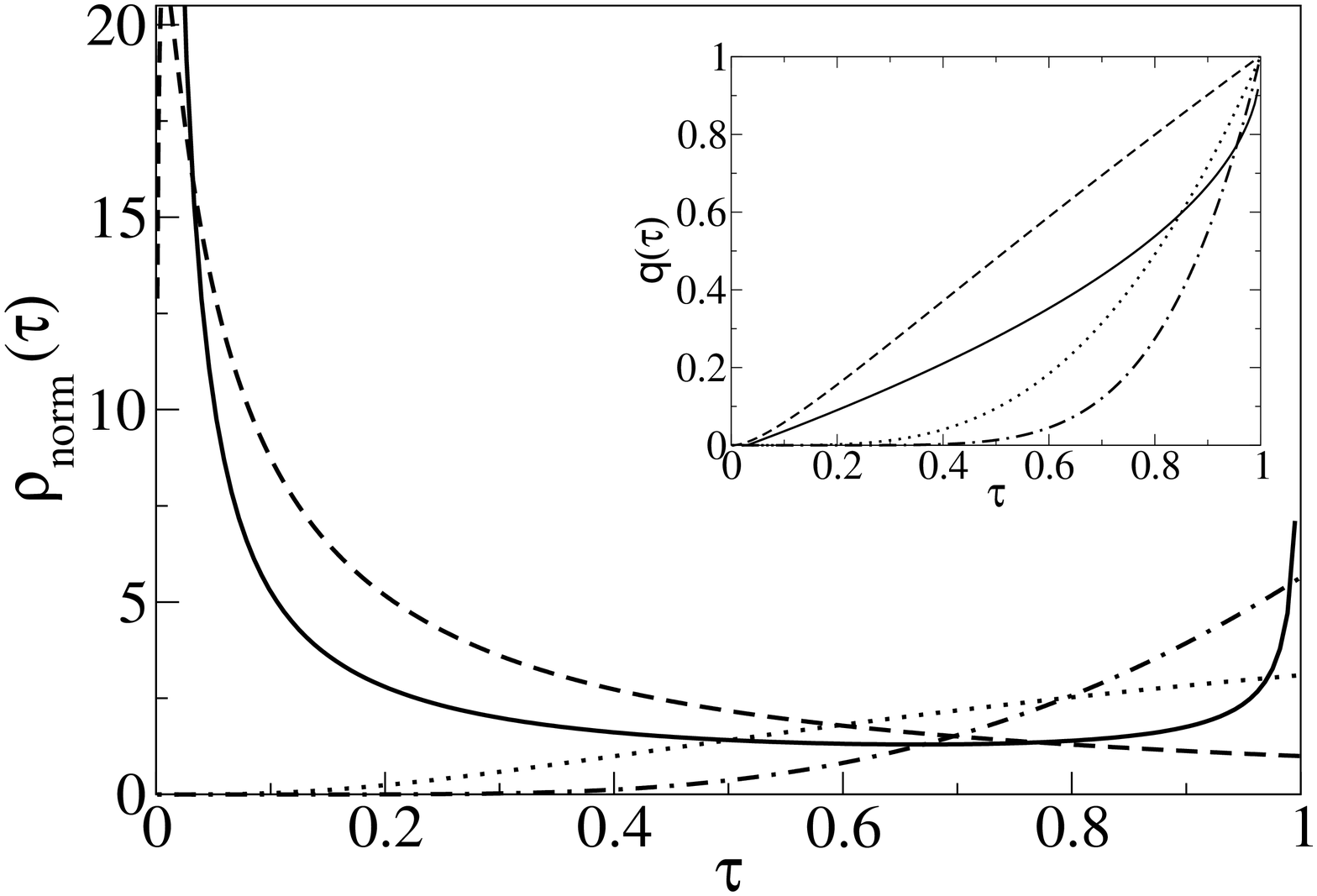}
\end{center}
\caption{\label{fig3} Main frame: Density $\rho_{\mathrm{norm}}(\tau)$ for 
$N_c=1$ at disorder parameter $2L/\xi$=2 (dashed  line), 2/5 (dotted line), 1/5 (dashed-dotted line) with average conductance  $\langle g \rangle=0.26,0.71, 0.83$, respectively. The solid line is computed from  Eq. (\ref{rho-metal}) and corresponds to the diffusive metallic regime. Inset: The cumulative function $q(\tau)$ given by Eq. (\ref{qoftau}) for 
the same values of the parameters $2L/\xi$ as in the main frame.}
\end{figure}

Performing numerically the integral in Eq. (\ref{rhotauNc1}),  we obtain the density $\rho_\mathrm{norm}(\tau)$ and the cumulative function $q(\tau)$  given by Eqs. (\ref{rhonorm}) and (\ref{qoftau}), respectively. We note that for the single channel case   $g = \tau $.
In Fig. \ref{fig3}, we show the evolution of $\rho_\mathrm{norm}(\tau)$ as the value of the disorder parameter $L/\xi$ is changed. 
As expected for $L > \xi$, small values of $\tau$, or conductance $g$, are favored, while as the ratio 
$L/\xi$ is decreased, the density is accumulated at values of $\tau $ near 1. As a reference, we have included the bimodal distribution (solid line) in Fig. \ref{fig3}. The function $q(\tau)$ is plotted in the inset of figure \ref{fig3} for the same values of  $L/\xi$ in the main frame. From Eq. (\ref{rho-metal}), in the metallic limit: $q_{\mathrm{met}}(\tau)= 1-\sqrt{1-\tau}$ (solid line in the inset of Fig. \ref{fig3}). In all the cases, $\rho_\mathrm{norm}(\tau)$ and $q(\tau)$ clearly exhibit discrepancies with respect to
the  diffusive metallic limit, in contrast to the results for the multichannel case, as we will see in the next subsection.

\subsection{\label{subsection_several}Several-channel case}

Let us now assume that the ballistic constriction can support several channels $N_c$.
Therefore from the expression for the density [Eq. (\ref{rhooftau})]   and using the join probability $p(\{x_i\})$ given by Eqs. (\ref{poflambda} - \ref{uandVmetal}), we can write the density of the transmission eigenvalues as 
\begin{eqnarray}
\label{rhomulti}
&&\rho(\tau) \propto \frac{e^{-\frac{\xi}{2L}x^2} \sqrt{x 
\sinh\left(2x \right)}}
{2\tau\sqrt{1-\tau}} 
\int \prod_i^{N_c} dx_i  \times \nonumber \\ & &  \exp\left[-\sum_{1<j}u(x,x_j) - \sum_{1<i<j}u(x_i,x_j)-\sum_{1<j} V_c(x_j) \right] ,\nonumber \\
\end{eqnarray}
where $x=\mathrm{arcsech}\sqrt\tau$,  $u(x_i,x_j)$ is given in Eq. (\ref{uandVmetal}), and 
\begin{equation}
V_c(x_i) = \frac{(N+1)}{2}\frac{l}{L}x_i^2-\frac{1}{2}\ln(x_i \sinh 2x_i) .
\end{equation}

We will see that  the density of the eigenvalues $\tau_n$ given by 
Eq. (\ref{rhomulti}) approximates to the diffusive metallic regime [Eq. (\ref{rho-metal})] even for a relative small number of channels $N_c$. 

We next present the results for $\rho_{\mathrm{norm}}(\tau)$ and $q(t)$ [Eqs. (\ref{rhonorm}) and (\ref{qoftau})] obtained from the above expression for $\rho(\tau)$ [Eq. (\ref{rhomulti})],  where the integrals were performed numerically. 
\begin{figure}
\includegraphics[width=1\columnwidth]{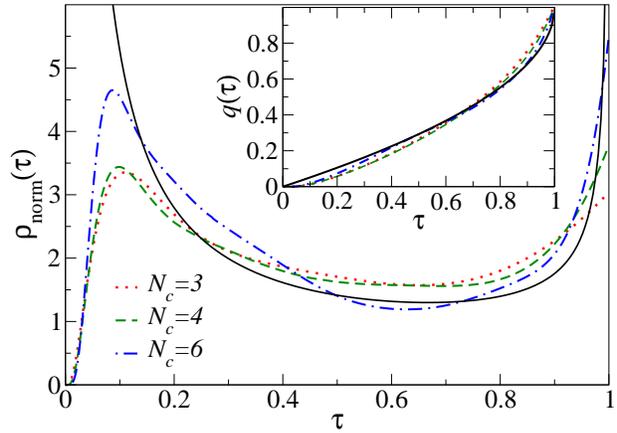}
\caption{\label{fig4}(Color on-line) Density $\rho_{\mathrm{norm}}(\tau)$ 
for $N_c= 3, 4,$ and 6 (dotted red, dashed green,  dashed-dotted blue lines, respectively). In the inset $q(\tau)$ is shown for the same values of $N_c$ in the main frame. For all cases, the average transmission $\langle \tau \rangle$ is approximately ($\langle \tau \rangle= 0.49 $), while the average conductance $ \langle g \rangle = 1.5, 2, 2.8$ for $N_c= 3, 4,$ and 6, respectively. The solid (black) line is the bimodal distribution in the metallic limit, Eq. (\ref{rho-metal}).}
\end{figure}

Let us start showing the evolution 
of the density of eigenvalues with the number of channels $N_c$. In figure \ref{fig4},  $\rho_{\mathrm{norm}}(\tau)$ and $q(t)$ are plotted for $N_c=3,4,$ and 6; these cases are chosen to have the same average value $\langle \tau \rangle$. We can observe  that as the number of channels is increased, $\rho_{\mathrm{norm}}(\tau)$ concentrates at small values of $\tau$ and near 1, i.e., the density evolves to the bimodal distribution plotted in solid line in the same Fig. \ref{fig4}.  In other words, as the conductance of the ballistic constriction increases, the distribution of the transmission eigenvalues approximates quickly the bimodal distribution. On the other hand, the cumulative function $q(\tau)$ (inset in Fig. \ref{fig4}) also comes close to
the result in the metallic limit $q_{\mathrm{met}}(\tau)$ (solid line). In fact, 
we note that the different curves for 
$q(\tau)$ might look closer to the metallic case $q_{\mathrm{met}}(\tau)$ than the corresponding cases for $\rho_{\mathrm{norm}}(\tau)$ to $\rho_{\mathrm{met}(\tau)}$ (the bimodal distribution). This is due to loss of details of the density $\rho_{\mathrm{norm}}(\tau)$ when this function is integrated [Eq. (\ref{qoftau})] to obtain $q(\tau)$.

\begin{figure}
\begin{center}
\includegraphics[width=1\columnwidth]{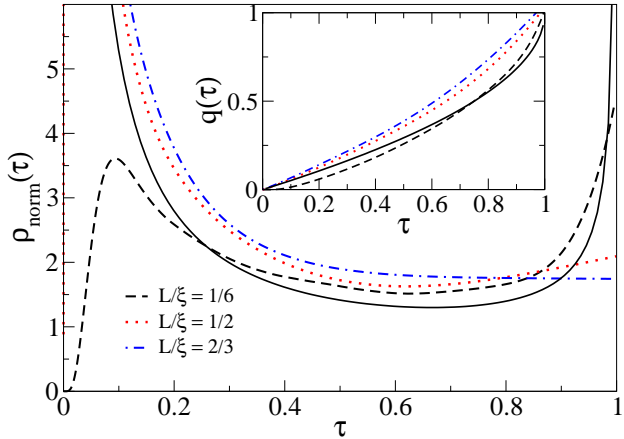}
\end{center}
\caption{\label{fig5}(Color-online) Evolution of the density $\rho_{\mathrm{norm}}(\tau)$ 
with the disorder parameter $L/\xi$. In the inset, $q(\tau)$ is plotted for the same values of $L/\xi$ in the main frame. In all the cases the number of channels is fixed to $N_c=5$. The solid (black) line is the bimodal density in the diffusive metallic regime. }
\end{figure}

Now we show the evolution of the density with the ratio $L/\xi$.  In figure \ref{fig5},  we
plot the results for $\rho_{\mathrm{norm}}(\tau)$ and $q(\tau)$ (inset) at three  different values of the disorder parameter $L/\xi$ for a fixed value of the number of channels  $N_c=5$. For the smallest value of $L/\xi$ (dashed line),   $\rho_{\mathrm{norm}}(\tau)$ shows a peak at $\tau$'s near 1, like the bimodal distribution (solid line); in fact, the density $\rho_{\mathrm{norm}}(\tau)$  looks similar to this metallic limit case, except at small values of $\tau$. As the ratio $L/\xi$ is increased, i.e., going to the insulating regime, the peak at small values of $\tau$ grows rapidly and $\rho_{\mathrm{norm}}(\tau)$ comes close to the bimodal distribution at small values of  $\tau$. An increase in peak height at small $\tau$'s, however, causes a detriment to the peak at $\tau$ near 1, as one might expect: roughly speaking, we can say that the evolution of $\rho_{\mathrm{norm}}(\tau)$ with the ratio $L/\xi$ described above is consequence of the normalization of the density function. Regardless of this behavior, for relatively small values of the ratio $L/\xi$, 
the main differences of $\rho_{\mathrm{norm}}(\tau)$ with respect to the diffusive metallic limit are  seen only at  small values of $\tau$, despite the small number of channels of the constriction.

\begin{figure}
\begin{center}
\includegraphics[width=0.9\columnwidth]{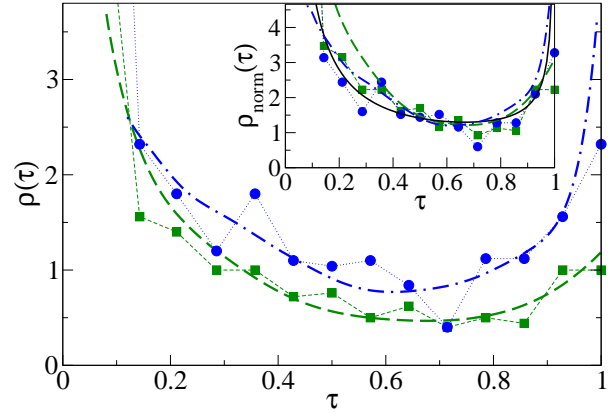}
\end{center}
\caption{\label{fig6}(Color on-line) Comparison between the experimental results  (solid symbols) for the density $\rho(\tau)$ for Pb nano-contacts and the theoretical calculations  (dashed and dashed-dotted lines). The conductances (in units of $e^2/h$) of the experimental data points are $g = 6$ (solid squares) and 10 (solid circles), also shown in the inset Fig. \ref{fig2}. The theoretical results are chosen to have  
average values $\langle \tau \rangle$ similar to the experimental ones, which we have estimated from the experimental data in the region ($0.14 <  \tau <1$), Fig. \ref{fig2}: $\langle \tau \rangle \approx$ 0.4 (solid squares), 0.5 (solid circles). The theoretical curves are normalized to the area of the experimental density in the region 
($0.14 <  \tau <1$). A good agreement is seen between the theoretical densities and experimental data points in the region where experimental results are reported. 
Inset: density  $\rho_{\mathrm{norm}}(\tau)$ for the same cases in the main frame. The 
solid (black) line is the bimodal density in the diffusive metallic limit.}
\end{figure}

Finally, we compare some of the experimental results of Refs. \cite{riquelme} and  \cite{memory} (Fig. \ref{fig2}) with  theoretical predictions. We point out that the theoretical model depends on the number of channels $N_c$ and the disorder parameter $L/\xi$ which we do not know from the experiments of Ref. \cite{riquelme}. Thus, in order to make the comparison, we have extracted the average transmission eigenvalue $\langle \tau \rangle$ from $\rho(\tau)$ plotted in the inset of Fig. \ref{fig2}. For the couple of chosen cases, solid squares and solid circles in Fig. \ref{fig2}, we have estimated: $\langle \tau \rangle  \approx 0.4$ and 0.5, respectively. We also note that these averages values of $\tau$  are calculated in the interval of $\tau$ ($0.14 < \tau < 1$ ), region where experimental data points have been reported.  From our calculations, the cases $N_c=6$ and 7 with $L/\xi =2/7$ and 1/8, respectively, give mean values  $\langle \tau \rangle$ similar to the experimental ones. In Fig. \ref{fig6} (main frame) we compare the results for $\rho(\tau)$: the dashed  line corresponds to $N_c=6$, while the dashed-dotted line is the result for $N_c=7$. In order to compare with the experimental results, we also have normalized the theoretical density to the area of the experimental density in the interval ($0.14 < \tau <1$). We can see that the experimental density 
is well described by theory. In the inset of figure \ref{fig6},  we also compare the normalized density $\rho_{\mathrm{norm}}(\tau)$ for the same cases in the main frame. The solid line in the inset corresponds to the bimodal distribution.  
\section{\label{section_conclusions}Summary}

We have studied the evolution of the transmission eigenvalue density of quantum point contacts with the disorder strength and 
the number of channels of the ballistic constriction. 
Our analysis of the transmission density is based 
on an approximation \cite{go-mu-wo,mu-wo-go} to the general solution of the DMPK equation and a mapping of the problem of transport through  a quantum wire with a ballistic constriction to the problem of transport through a wire without the constriction \cite{carlo-melsen} i.e., the latter problem  is described by the known solution of the DMPK equation. We have shown that the density of the transmission eigenvalues comes close to the known bimodal distribution in the diffusive metallic regime even for a small number of channels supported by the constriction of the quantum point contact. This is in agreement with the unexpected result pointed out in Ref. \cite{riquelme}, where some statistical properties of the transmission eigenvalues for Pb nano-contacts were analyzed by using MCBJ techniques. In order to apply the DMPK approach to those experiments, we consider that the neck formed during the elongation process in a MCBJ device is a ballistic constriction which is in between two wide disordered regions. We do not have experimental information of each parameter that enters in the theoretical model, but by comparing theoretical and experimental results with similar average transmission values $\langle \tau \rangle$, we have seen that the analyzed experimental densities are well described by the corresponding
theoretical densities. 
The trend of the transmission density seen in  Pb contacts experiments as the number of channels is changed is also in agreement with the theoretical results presented here when the number of channels of the ballistic constriction is varied.

\begin{acknowledgement}
I thank J. J. Riquelme for providing his Memory of Research as well as G. Rubio-Bollinger. I also acknowledge useful correspondence with A. Levy Yeyati and C. Urbina. This work was supported by the Ministerio de Educaci\'on y Ciencia through the Ram\'on y Cajal Program and the Fondo Social Europeo.
\end{acknowledgement}


\begin{thebibliography}{40}
\bibitem{wees}
B. J. van Wees, H. van Houten, C. W. J. Beenakker, J. G. Williamson, L. P. Kouwenhoven, D. van der Marel, and C.T. Foxon, Phys. Rev. Lett. \textbf{60}, (1988) 848.

\bibitem{wharam}
D A Wharam, T J Thornton, R Newbury, M Pepper, H Ahmed, J E F Frost, D G Hasko, D C Peacock, D A Ritchie and G A C Jones, J. Phys. C \textbf{21}, (1988) L209.
\bibitem{agrait} 
N. Agra\"it,A. Levy Yeyati, and Jan M. van Ruitenbeek, Phys. Rep. \textbf{377}, (2003) 81.

\bibitem{carlo-review}
C. W. J. Beenakker,
Rev. Mod. Phys. \textbf{69}, (1997) 731.

\bibitem{maslov} D. L. Maslov, C. Barnes, and G. Kirczenow, Phys. Rev. Lett. \textbf{70}, (1993) 1984.

\bibitem{carlo-melsen}
C. W. J. Beenakker and J. A. Melsen, Phys. Rev. B \textbf{50}, (1994) 2450.


\bibitem{campagnano}
G. Campaganano, O. N. Jouravlev, Ya. M. Blanter, and Yu. V. Nazarov, Phys. Rev. B \textbf{69}, (2004) 235319.

\bibitem{elke} E. Scheer, et. al. Nature \textbf{394}, (1998) 154.

\bibitem{smit} R. H. M. Smit, C. Untiedt, G. Rubio-Bollinger, R. C. Segers, and J. M. van Ruitenbeek, Phys. Rev. Lett. \textbf{91}, (2003) 076805-1.


\bibitem{riquelme} J. J. Riquelme, L. de la Vega, A. Levy Yeyati, N. Agra\"it, A. Martin-Rodero, and G. Rubio-Bollinger, Europhys. Lett., \textbf{70}, (2005) 663.

\bibitem{memory} J. J. Riquelme, Memoria de Investigaci\'on, Universidad Aut\'onoma de Madrid.

\bibitem{pier-book} P. A. Mello and N. Kumar, {\it Quantum Transport in
  Mesoscopic Systems. Complexity and statistical fluctuations}, Oxford
  University Press, Oxford, 2004.

\bibitem{dorokhov} O. N. Dorokhov, JETP Lett.  \textbf{36}, (1982) 318.

\bibitem{rejaei} 
C. W. J. Beenakker and  B. Rejaei, Phys. Rev. Lett. \textbf{71}, (1993) 3689.


\bibitem{go-mu-wo}
Victor A. Gopar, K. A. Muttalib, Peter W\"olfle, Phys. Rev. B, \textbf{66}, (2002) 174204.

\bibitem{mu-wo-go}
K. A. Muttalib, Peter W\"olfle, Victor A. Gopar, Ann. Phys., \textbf{308}, (2003) 156.


\bibitem{abrikosov} A. A. Abrikosov, Solid State Commun. {\bf 37}, (1981) 997.

\bibitem{caselle} M. Caselle, Phys. Rev. Lett. \textbf{74}, (1995) 2776.

\bibitem{pier-pichard} P. A. Mello and J.-L. Pichard, Phys. Rev. B \textbf{40}, (1989) 5276.

\bibitem{data} The data points to produce Figure 2 have been extracted from Fig. 2 in Ref. \cite{riquelme} and Fig. 8 in Ref. \cite{memory}.

\end{thebibliography}
\end{document}